\documentclass[fleqn]{article}

\usepackage{amsfonts}
\usepackage{amsmath}
\usepackage[amsmath,thmmarks]{ntheorem}
\usepackage{url}
\usepackage{tikz}
\usetikzlibrary{quantikz}
\usepackage{xspace}
\usepackage[latin1]{inputenc}
\usepackage{paralist}

\newtheorem{theorem}{Theorem}

{\theorembodyfont{\upshape}
  \theoremsymbol{\ensuremath{\Diamond}}
  \newtheorem{definition}[theorem]{Definition}
  
  \theoremsymbol{}
  \newtheorem{example}[theorem]{Example}
  \theoremstyle{nonumberplain}
  \theoremsymbol{}
\theoremseparator{.}
  
}
\theoremheaderfont{\sc}\theorembodyfont{\upshape}
\theoremstyle{nonumberplain}
\theoremseparator{.}
\theoremsymbol{\rule{1ex}{1ex}}

\newcommand{\isdef}{\stackrel{\mbox{\tiny def}}{=}}

\newcommand{\set}[1]{\left\{#1\right\}}
\newcommand{\tuple}[1]{\langle #1 \rangle}
\newcommand{\card}[1]{\left\vert#1\right\vert}
\newcommand{\norm}[1]{\left\lVert#1\right\rVert}
\newcommand{\setof}[2]{\left\{#1\,\middle|\:#2\right\}}

\newcommand{\dN}{\mathbb{N}}

\newcommand{\dR}{\mathbb{R}}
\newcommand{\dB}{\mathbb{B}}

\newcommand{\dH}{\mathcal{H}}
\newcommand{\dC}{\mathbb{C}}
\newcommand{\dE}{\mathbb{E}}

\newcommand{\calA}{\mathcal{A}}
\newcommand{\calM}{\mathcal{M}}
\newcommand{\calN}{\mathcal{N}}
\newcommand{\calC}{\mathcal{C}}
\newcommand{\calE}{\mathcal{E}}
\newcommand{\keyw}[1]{\textbf{#1}}
\newcommand{\interv}[2]{[#1,#2]}

\newcommand{\inner}[2]{\langle #1|#2\rangle}
\newcommand{\outerp}[2]{\ket{#1}\!\bra{#2}}
\newcommand{\conj}[1]{\overline{#1}}
\newcommand{\adj}[1]{#1^\dagger}
\newcommand{\vzero}{\mathbf{0}}
\newcommand{\idty}{\mathbf{I}}
\newcommand{\trc}[1]{\mathrm{Tr}(#1)}
\newcommand{\cmat}[2]{\dC^{#1\times #2}}

\newcommand{\lmspace}{\vspace{0.5em}}
\newcommand{\aindent}{\quad}
\newcommand{\sindent}{}

\newcommand{\lhv}{\texttt{lhv}}
\newcommand{\msr}{\texttt{measure}}
\newcommand{\pbspace}{\texttt{prob-space}}
\newcommand{\bmeas}{\texttt{borel-measurable}}
\newcommand{\posrv}{\texttt{pos-rv}}
\newcommand{\psrv}{\texttt{prv-sum}}
\newcommand{\fcmat}{\texttt{fixed-carrier-mat}}
\newcommand{\dimr}{\texttt{dimR}}
\newcommand{\dimc}{\texttt{dimC}}
\newcommand{\fcmats}{\texttt{fc-mats}}
\newcommand{\summat}{\texttt{sum-mat}}
\newcommand{\sumwith}{\texttt{sum-with}}
\newcommand{\isaset}{\texttt{set}}
\newcommand{\cpxsqmat}{\texttt{cpx-sq-mat}}

\begin{document}
	\title{Quantum projective measurements and the CHSH inequality in Isabelle/HOL}

\author{Mnacho Echenim and Mehdi Mhalla\\ {Universit\'e Grenoble Alpes, Grenoble INP}\\ {CNRS, LIG, F-38000 Grenoble, France}
}

\date{March 2021}


\maketitle
\begin{abstract}
We present a formalization in Isabelle/HOL of quantum projective measurements, a class of measurements involving orthogonal projectors that is frequently used in quantum computing. We also formalize the CHSH inequality, a result that holds on arbitrary probability spaces, which can used to disprove the existence of a local hidden-variable  theory for quantum mechanics.
\end{abstract}

\section{Introduction}

One of the (many) counterintuitive aspects of quantum mechanics involves the measurement postulates, which state that:
\begin{itemize}
	\item Measurements of equally prepared systems such as photons or electrons do not always output the same value, which implies that it is only possible to make statistical predictions on properties of these objects;
	\item Once the measurement of an object has produced an outcome $\lambda$, all subsequent measurements of the same object also produce the outcome $\lambda$, a phenomenon known as the \emph{collapse of the wave function}.
\end{itemize}
These postulates are now widely accepted, but this has not always been the case, as evidenced by the well known EPR paradox, named after Einstein, Podolsky and Rosen, in a paper published in 1935, suggesting that quantum mechanics was an incomplete theory 
 \cite{EPR}. In their thought experiment, the authors considered two entangled particles that are separated. When one of them is measured, the collapse of the wave function guarantees that the measure of the other particle will produce the same result. Because instantaneous transmission of information is impossible, the authors viewed this phenomenon as evidence that there existed hidden variables that predetermined the measurement outcomes but were not accounted for by quantum mechanics. The search for so-called \emph{local hidden-variable theories} for quantum mechanics came to a halt in 1964 when Bell proved \cite{Bell64} that these theories entailed upper-bounds on the correlations of measure outcomes of entangled particles, and that these upper-bounds are violated by the correlations predicted by quantum mechanics; a violation that has since then been experimentally confirmed \cite{aspect}.

In this paper we present the formalization in Isabelle/HOL of \emph{projective} measurements, also known as von Neumann measurement, along with the related notion of \emph{observables}. Projective measurements involve complete sets of orthogonal projectors that are used to compute the probabilities of measurement outcomes and to determine the \emph{state collapse} after the measurement. Although there are more general forms of measurements in quantum mechanics, projective measurements are quite common: for example, textbooks on quantum mechanics frequently mention ``measuring a state in a given orthonormal basis''; an operation that means performing a projective measurement with the projectors on the elements of the orthonormal basis. These measurements are especially used in quantum computing and quantum information, and are thus of a particular interest to computer scientists.

Next we formalize the \emph{CHSH inequality}, named after Clauser, Horne, Shimony and Holt \cite{CHSH}. This inequality involving expectations of random variables can be used to prove Bell's theorem that quantum mechanics cannot be formalized in a probabilistic setting involving local hidden variables, thus showing the intrinsically statistical nature of quantum mechanics.

\paragraph{Related work.}
There are several lines of ongoing research on the use of formal tools for the analysis of quantum algorithms and protocols, such as an extension of attack trees with probabilities \cite{Kammuller19}, or the development of dedicated quantum Hoare logics \cite{Unruh_2019,liu19}. Approaches closer to ours involve the formalization of quantum notions and algorithms in proof assistants including Coq \cite{Boender_2015,Coq_Quantum} and Isabelle \cite{isa-dirac}. In this paper we extend the effort started in \cite{isa-dirac} in two ways. First, our formalization is based on so-called \emph{density operators} rather than pure quantum states. Although both notions are equivalent, density operators are a more convenient way of representing quantum systems that are in a mixed state, i.e., in one of several pure quantum states with associated probabilities that sum to 1. They also permit to obtain more general and natural statements on quantum mechanics. Second, although notions related to measurements are formalized in \cite{isa-dirac}, these are specific and in particular, only involve measurements in the standard basis. We develop the full formalization of projective measurements and observables in this paper. To the best of our knowledge, no such formalization is available in any proof assistant.

Our formalization is available on the \emph{Archive of Formal Proofs}, at the address \url{https://www.isa-afp.org/entries/Projective_Measurements.html}. It is decomposed in three parts: \begin{inparaenum}[(i)]\item A formalization of necessary notions from Linear Algebra. We heavily rely on the results developed in \cite{liu19}, especially those involving complex matrices and the decomposition of Hermitian matrices for this part, as well as the \emph{types-to-sets} transfer tool \cite{types-sets} to use general results on types in our setting. \item The formalization of projective measurements and their relationship to observables. In particular we show how to construct a projective measurement starting from an observable and how to recover the observable starting from the projective measurement. \item The formalization of the CHSH inequality. Along with the proof of the inequality on an arbitrary probability space, we prove that assuming the existence of a local hidden variable to explain the outcome of measurements leads to a contradiction. \end{inparaenum}


\section{Preliminaries}

We review the formalization of probability theory in Isabelle, as well as basic notions from linear algebra and quantum mechanics. A more detailed presentation on this last topic can be found in, e.g., \cite{mathQuantum, nielsen-book}. As we will only formalize quantum notions in finite dimension, we present the standard general definitions and illustrate them in the finite dimensional case.

\subsection{Probability theory in Isabelle}

We begin by briefly presenting the syntax of the interactive theorem prover Isabelle/HOL; this tool can be downloaded at \url{https://isabelle.in.tum.de/}, along with tutorials and documentations. Additional material on Isabelle can be found in \cite{ConcreteSemantics}. This prover is based on higher-order logic; terms are built using types that can be: 
\begin{itemize}
	\item simple types, denoted with the Greek letters $\alpha, \beta,\ldots$
	\item types obtained from type constructors, represented in postfix notation (e.g. the type $\alpha$ \texttt{set} denotes the type of sets containing elements of type $\alpha$), or in infix notation (e.g., the type $\alpha\rightarrow \beta$ denotes the type of total functions from $\alpha$ to $\beta$).
\end{itemize}
Functions are curried, and function application is written without parentheses. Anonymous functions are represented with the lambda notation: the function $x\mapsto t$ is denoted by $\lambda x.\,t$. We will use  mathematical notations for standard terms; for example, the set of reals will be denoted by $\dR$ and the set of booleans by $\dB$. The application of function $f$ to argument $x$ may be written $f\ x$, $f(x)$ or $f_x$ for readability.

\newcommand{\carrier}{\texttt{carrier-mat}}
\newcommand{\gmat}{\texttt{mat}}

A recent tool \cite{types-sets} permits to transfer type-based statements which hold on an entire type universe to their set-based counterparts. This tool is particularly useful to apply lemmata in contexts where the assumptions only hold on a strict subset of the considered universe. For example generalized summation over sets is defined for types that represent abelian semigroups with a neutral element. Such an algebraic structure is straightforward to define on any set of matrices that all have the same  dimensions. In our formalization, we define a \emph{locale} \cite{Ballarin14} for such a set of matrices in which the number of rows and columns is fixed:
\[\begin{array}{l}
	\keyw{locale}\ \fcmat\ =\  \keyw{fixes}\ \fcmats\ \dimr\ \dimc\\
	\quad\keyw{assumes}\ \fcmats\ =\ \carrier\ \dimr\ \dimc 
\end{array}\]
The set of matrices with $n$ rows and $m$ columns is represented in Isabelle by $\carrier\ n\ m$, 
all matrices in {\fcmats} thus admit $\dimr$ rows and $\dimc$ columns. After proving that $\fcmats$ along with the standard addition on matrices and the matrices with $\dimr$ rows and $\dimc$ columns consisting only of zeroes is an abelian semigroup with a neutral element, we can define a generalized summation of matrices on this locale:
\[\begin{array}{lcl}
	\summat & :: & (\alpha \rightarrow \beta\ \gmat) \rightarrow \alpha\ \isaset\ \rightarrow \beta\ \gmat\\
	\summat\ \calA\ I &= & \sumwith\ (+)\ \vzero\ \calA\ I
\end{array}\]
The types-to-sets transfer tool then permits with no effort to transfer theorems that hold on abelian semigroups, and especially those involving sums, to this locale.

A large part of the formalization of measure and probability theory in Isabelle was carried out in \cite{hoelzl2012thesis} and is  included in Isabelle's distribution.  We briefly recap some of the notions that will be used throughout the paper and the way they are formalized in Isabelle. We assume the reader has knowledge of fundamental concepts of measure and probability theory; any missing notions can be found in 
\cite{Durrett} for example. 
Probability spaces are particular \emph{measure spaces}.  A measure space over a set $\Omega$ consists of a function $\mu$ that associates a nonnegative number {or $+\infty$} to  some subsets of $\Omega$. {The subsets of $\Omega$ that can be measured are closed under  complement and countable unions and make up a \emph{$\sigma$-algebra}.} In Isabelle the measure type with elements of type $\alpha$ is denoted by $\alpha\ \msr$.
\newcommand{\measurable}{\texttt{measurable}}

A function between two measurable spaces is \emph{measurable} if the preimage of every measurable set is measurable. 
In Isabelle, sets of measurable functions are defined as follow:
\[\begin{array}{lcl}
	\measurable & :: & \alpha\,\msr \rightarrow \beta\,\msr\rightarrow \left(\alpha \rightarrow \beta\right) \isaset\\
	\measurable\ \calM\ \calN\ \mu & =&
	\setof{f: \Omega_\calM \rightarrow \Omega_\calN}{\forall A\in \calA_\calN.\, f^{-1}(A) \cap \Omega_\calM \in \calA_M}
\end{array}\]
Measurable functions that map the elements of a measurable space into real numbers, such as  random variables which are defined below, are measurable on Borel sets:

\[\begin{array}{l}
	\keyw{abbreviation}\ \bmeas\ \calM\ \equiv\  \measurable\ \calM\ \texttt{borel}  
\end{array}\]

\newcommand{\probspace}{\texttt{prob-space}}
\newcommand{\finitemeasure}{\texttt{finite-measure}}
\newcommand{\AEv}{\textsc{AE}}

Probability measures are measure spaces on which the measure of $\Omega$ is finite and equal to 
$1$. In Isabelle, they are defined in a {locale}.

\[\begin{array}{l}
	\keyw{locale}\ \probspace =  \finitemeasure\ +\keyw{assumes}\ \mu_\calM(\Omega_\calM)\ =\ 1 
\end{array}\]
A \emph{random variable} on a probability space $\calM$ is a measurable function with domain $\Omega_\calM$. The average value of a random variable $f$ is called its \emph{expectation}, it is denoted by\footnote{The superscript is omitted when there is no confusion.} $\dE^\calM[f]$, and defined by $\dE^\calM[f] \isdef \int_{\Omega_\calM}f\mathrm{d}\mu_\calM$.

In what follows, we will consider properties that hold \emph{almost surely} (or \emph{almost everywhere}), {i.e.,} are such that the elements for which they do not hold reside within a set of measure $0$:
\begin{align*}
	\keyw{lemma}\ \textsc{AE-iff}& :\\\
	(\AEv_\calM\,x.\ P\ x)& \Leftrightarrow (\exists N\in \calA_\calM.\, \mu_\calM(N) = 0 \wedge \setof{x}{\neg P\ x}\subseteq N)
\end{align*}
We will formalize results involving local hidden-variable hypotheses under the more general assumption that properties hold almost everywhere, rather than on the entire probability space under consideration.

\subsection{On linear algebra}

\newcommand{\cpxmat}{\texttt{complex }\gmat}
\newcommand{\spct}[1]{\mathrm{spct}(#1)}

We recap the core notions from linear algebra that will be used in our formalism. A more detailed treatment can be found, e.g., in \cite{linalg}.
A \emph{Hilbert space} $\dH$ is a complete vector space over the field of complex numbers, equipped with an \emph{inner product} $\inner{\cdot}{\cdot}: \dH \times \dH \rightarrow \dC$, i.e., a function such that for all $\varphi, \psi\in \dH$, $\inner{\varphi}{\psi} = \conj{\inner{\psi}{\varphi}}$, $\inner{\varphi}{\varphi} \geq 0$ and $\inner{\varphi}{\varphi} = 0$ iff $\varphi = \vzero$. The \emph{norm} induced by the inner product is defined by $\norm{\varphi} \isdef \sqrt{\inner{\varphi}{\varphi}}$, and $\varphi$ is \emph{normalized} if $\norm{\varphi} = 1$. The elements of a Hilbert space of dimension $n$ are represented as column vectors, and the inner product of $\varphi$ and $\psi$ is $\inner{\varphi}{\psi} = \sum_{i = 1}^n \conj{\varphi_i}\psi_i$.

An \emph{operator} is a linear map on a Hilbert space. We denote by $\idty$ the identity operator; we may also write $\idty_n$ to specify that the considered Hilbert space is of dimension $n$. If $A$ is an operator on $\dH$, then the operator $B$ such that for all $\varphi, \psi \in \dH$, $\inner{\varphi}{A\psi} = \inner{B\varphi}{\psi}$, is called the \emph{adjoint of $A$}, and denoted by $\adj{A}$. If $A = \adj{A}$ then we say that $A$ is \emph{self-adjoint} or \emph{Hermitian},  and if $A\adj{A} = \adj{A}A = \idty$, then we say that $A$ is \emph{unitary}. We identify operators with their matrix representation.  The set of complex matrices is represented in Isabelle by \cpxmat. For convenience, we denote by $\cmat{n}{m}$ the set of complex matrix with $n$ rows and $m$ columns and by $\vzero_{n,m}$ the matrix in $\cmat{n}{m}$ containing only zeroes. When there is no ambiguity, we will write $\vzero$ instead of $\vzero_{n,m}$. The adjoint of the (square) matrix $A$ satisfies $(\adj{A})_{i,j} = \conj{A_{j,i}}$, and if $A$ is self-adjoint then we say that $A$ is a \emph{Hermitian matrix}. The \emph{trace} of a square matrix is the sum of its diagonal elements: if $A\in \cmat{n}{n}$ then $\trc{A} = \sum_{i=1}^n A_{i,i}$. An operator $A$ is a \emph{projector} if $A^2 = A$, and $A$ is an \emph{orthogonal projection} if $\adj{A} = A$. We say that $A$ is \emph{positive} if, for all $\varphi \in \dH$, we have $\inner{\varphi}{A\varphi} \geq 0$.

We say that the vector $\varphi$ is an \emph{eigenvector} of operator $A$ is there exists $\lambda\in \dC$ such that $A\varphi = \lambda \varphi$. In this case, we say that $\lambda$ is an \emph{eigenvalue} of $A$. The set of eigenvalues of $A$ is called the \emph{spectrum} of $A$ and denoted by $\spct{A}$. When $A$ is Hermitian, all its eigenvalues are necessarily real and it is possible to associate every eigenvalue $\lambda \in \spct{A}$ with a projector $P_\lambda$ such that
\[P_\lambda\cdot P_{\lambda'} = \vzero\ \text{if}\ \lambda \neq \lambda',\ \sum_{\lambda \in \spct{A}}P_\lambda = \idty\ \text{and}\ A\ =\ \sum_{\lambda \in \spct{A}}\lambda P_\lambda.\]
\newcommand{\hermitian}{\texttt{hermitian}}
\newcommand{\dimrow}{\texttt{dim-row}}
\newcommand{\unitary}{\texttt{unitary}}

The \emph{tensor product} of two vector spaces $U$ and $V$ is denoted by $U\otimes V$. We also consider the tensor product of vectors $u\in U$ and $v\in V$, and denote it by $u\otimes v$; similarly, for matrices $A$ and $B$, we denote by $A\otimes B$ their tensor product. Two properties of interest in this context are the following:
\[\begin{array}{l}
	\keyw{lemma}\ \textsc{tensor-mat-hermitian}:\\
	\aindent \keyw{assumes}\ A \in \carrier\ n\ n\ \keyw{and}\ B\in \carrier\ n'\ n'\\
	\aindent \keyw{and}\ n > 0\ \keyw{and}\ n' > 0\\
	\aindent \keyw{and}\ \hermitian\ A\ \keyw{and}\ \hermitian\ B\\
	\sindent \keyw{shows}\ \hermitian\ A\otimes B\lmspace\\
	\keyw{lemma}\ \textsc{tensor-mat-unitary}:\\
	\aindent \keyw{assumes}\ \dimrow\ A > 0\ \keyw{and}\ \dimrow\ B > 0\\
	\aindent \keyw{and}\ \unitary\ A\ \keyw{and}\ \unitary\ B\\
	\sindent \keyw{shows}\ \unitary\ A\otimes B\\
\end{array}\]

\newcommand{\rankproj}{\texttt{rank-1-proj}}
\newcommand{\adjoint}{\texttt{adjoint}}
\newcommand{\projector}{\texttt{projector}}
\newcommand{\trace}{\texttt{trace}}

It is standard in quantum mechanics to represent the elements of $\dH$ using the Dirac notation: these elements are called \emph{ket-vectors}, and are denoted by $\ket{u}$. In what follows, we will represent the tensor product $\ket{u}\otimes \ket{v}$ by $\ket{uv}$. We also consider the dual notion of \emph{bra-vectors}, denoted by $\bra{u}$. Formally, a bra-vector is the linear map that maps the vector $\ket{v}$ to the complex number $\inner{u}{v}$. In the finite-dimensional setting, if \[\ket{u} = \begin{pmatrix}
	u_{1} \\
	u_{2} \\
	\vdots \\
	u_{n}
\end{pmatrix}\]
then $\bra{u} = (\conj{u_1}, \conj{u_2}, \ldots, \conj{u_n})$. We will write $\bra{uv}$ instead of $\bra{u}\otimes \bra{v}$, and by a slight abuse of notation, we will identify the application of $\bra{u}$ to $\ket{v}$ with the inner product $\inner{u}{v}$.
We define the \emph{outer product} of two vectors, denoted by $\outerp{u}{v}$, as the linear map such that, for all $\ket{v'} \in \dH$, $(\outerp{u}{v})\ket{v'} = \ket{u}\cdot\inner{v}{v'} = \inner{v}{v'}\cdot \ket{u}$. In the finite-dimensional setting, this outer product is represented by the matrix $M$, where $M_{i,j} = u_i\conj{v_j}$.	When $\ket{v}$ is normalized, the outer product $\outerp{v}{v}$ is called the \emph{rank-1 projection on $v$}. It is in fact an orthogonal projection with trace 1:
\[\begin{array}{l}
	\keyw{lemma}\ \textsc{rank-1-proj-adjoint}:\\
	\sindent \keyw{shows}\ \adjoint\ (\rankproj\ v)\ = \rankproj\ v\lmspace\\
	\keyw{lemma}\ \textsc{rank-1-proj-unitary}:\\
	\aindent \keyw{assumes}\ \norm{v} = 1\\
	\sindent \keyw{shows}\ \projector\ (\rankproj\ v)\lmspace\\
	\keyw{lemma}\ \textsc{rank-1-proj-trace}:\\
	\aindent \keyw{assumes}\ \norm{v} = 1\\
	\sindent \keyw{shows}\ \trace\ (\rankproj\ v)\ = 1
\end{array}\]

Many introductory textbooks on quantum computing present quantum postulates on states, i.e., normalized vectors on the underlying Hilbert space. The more general statement of these postulates involves so-called \emph{density operators}, which are represented by positive matrices of trace 1. This notion has already been formalized in \cite{liu19}: 
\newcommand{\densop}{\texttt{density-operator}}
\[\begin{array}{lcl}
	\densop & :: & \cpxmat \rightarrow \dB\\
	\densop\ \rho &\equiv & \texttt{positive}\ \rho\ \wedge\ \trace\ \rho\ = 1
\end{array}\]

\begin{example}
	Given a Hilbert space $\dH$, consider a set of normalized vectors  $\ket{u_1}, \ldots \ket{u_n}$ and assume that for $i\in \interv{1}{n}$, $p_i\geq 0$ and $\sum_{i = 1}^n p_i = 1$. Then 
	\[\rho \ \isdef\ \sum_{i=1}^n p_i \outerp{u_i}{u_i}\]
	is a density operator (lemma \textsc{rank-1-proj-sum-density}). In quantum terminology, when $n=1$, the matrix $\rho$ is referred to as a \emph{pure state} and otherwise, to a \emph{mixed state}.
\end{example}

A density operator that will be used in the formalization of the measurement postulate is the so-called \emph{maximally mixed state}. It is formalized as follows in Isabelle:
\newcommand{\maxmix}{\texttt{max-mix-density}}

\[\begin{array}{lcl}
	\maxmix & :: & \dN \rightarrow \cpxmat\\
	\maxmix\ n &= & \frac{1}{n}\cdot \idty_n
\end{array}\]
Intuitively, this density operator is the one with the maximal von Neumann entropy; in other words, its spectrum admits the maximum Shannon entropy.


All the matrices we consider in what follows are nontrivial complex square matrices, we thus define a locale to work in this context:
\[\begin{array}{l}
	\keyw{locale}\ \cpxsqmat\ =\  \fcmat\ (\fcmats:: \cpxmat\ \isaset)\, +\\
	\quad\keyw{assumes}\ \dimr = \dimc\ \keyw{and}\ \dimr > 0
\end{array}\]

\subsection{Quantum mechanics postulates}

We present the postulates of quantum mechanics that are used in quantum computation and information, following \cite{nielsen-book}. 
\begin{description}
	\item[State postulate] Associated to an isolated physical system is a Hilbert space, which is referred to as the \emph{state space}. The system itself is completely described by its density operator; we thus identify the state of a system with its density operator.
	\item[Evolution postulate] The evolution of a closed quantum system is described by a unitary transformation: the state $\rho$ at time $t$ of the system is related to the state $\rho'$ at time $t'$ of the system by a unitary operator $U\isdef U(t,t')$, by the equation $\rho' = U\rho \adj{U}$.
	\item[Measurement postulate] Quantum measurements are described by collections of so-called \emph{measurement operators}. These collections are of the form $\setof{M_\alpha}{\alpha\in I}$, where $M_\alpha$ is the matrix associated to the measurement outcome $\alpha$, and they satisfy the completeness equation:
	\[\sum_{\alpha\in I}\adj{M_\alpha}M_\alpha\ =\ \idty.\]
	When the state $\rho$ of a quantum system is measured with the collection $\setof{M_\alpha}{\alpha\in I}$, the probability that the outcome $\alpha$ occurs is $\trc{\adj{M_\alpha}M_\alpha\rho}$, and the state after the measurement \emph{collapses} into 
	\[\rho'\ \isdef\ \frac{M_\alpha\rho\adj{M_\alpha}}{\trc{\adj{M_\alpha}M_\alpha\rho}}.\]
	\item[Composite postulate] The state space of a composite physical system is the tensor product of the state spaces of the component physical systems. If the system consists of $n$ individual systems and system $i$ is prepared in state $\rho_i$ for $i\in \interv{1}{n}$, then the joint state of the composite system is $\rho_1\otimes \rho_2\otimes \cdots \otimes \rho_n$.
\end{description}

\begin{example}
	The simplest quantum system is described by a two-dimensional Hilbert space. It is standard to note an orthonormal basis for such a vector space as $\set{\ket{0}, \ket{1}}$; the two elements of this orthonormal basis are called the \emph{computational basis states}. In the density operator terminology, a \emph{qubit} is a state of the form $\outerp{\varphi}{\varphi}$, where $\ket{\varphi} = a \ket{0} + b \ket{1}$, for some $a,b\in \dC$ such that $\card{a}^2 + \card{b}^2 = 1$. 
\end{example}

In Section \ref{sec:chsh}, we will consider composite systems involving two qubits. Consider two physical systems $A$ and $B$, to which are associated the Hilbert spaces $\dH_A$ and $\dH_B$. Although the state space of the composite system consisting of $A$ and $B$ is simply $\dH_A\otimes \dH_B$, properties of measurements involving this composite system can be quite counterintuitive, because of the notion of entanglement:
\begin{definition}
	A state $\rho$ is \emph{separable} if it can be written as
	\[\rho\ =\ \sum_{i = 1}^n p_i \rho^i_A\otimes \rho^i_B,\]
	where for $i \in \interv{1}{n}$, $0 \leq p_i\leq 1$ and $\sum_{i=1}^n p_i = 1$. Otherwise, it is \emph{entangled}.
\end{definition}

For systems involving two qubits, we have the \emph{Bell states} are examples of entangled pure states:
\begin{eqnarray*}
	\ket{\Phi^+} & \isdef & \frac{1}{\sqrt{2}}\left(\ket{00} + \ket{11}\right)\\
	\ket{\Phi^-} & \isdef & \frac{1}{\sqrt{2}}\left(\ket{00} - \ket{11}\right)\\
	\ket{\Psi^+} & \isdef & \frac{1}{\sqrt{2}}\left(\ket{01} + \ket{10}\right)\\	
	\ket{\Psi^-} & \isdef & \frac{1}{\sqrt{2}}\left(\ket{01} - \ket{10}\right)
\end{eqnarray*}
The Bell state $\ket{\Psi^-}$ will be used in Section \ref{sec:chsh} to contradict the local hidden variable hypothesis.

\section{Projective measurements}

\subsection{The projective measurement postulate}

A notion related to collections of measurement operations is that of \emph{observables}. Intuitively, an observable represents a physically measurable quantity of a quantum system, such as the spin of an electron or the polarization of a photon. Formally, observables are represented by Hermitian operators; this is the approach that was used by von Neumann in his axiomatic treatment of quantum mechanics. This is why it is common to see the Measurement postulate stated as follows:
\begin{description}
	\item[Projective measurement postulate] A projective measurement is described by an observable, which is a Hermitian operator $M$ on the state space of the observed system. The measurement outcomes for an observable are its eigenvalues. Given the spectral decomposition 
	\[M\ =\ \sum_{\lambda \in \spct{M}} \lambda\cdot P_\lambda,\]
	where $P_\lambda$ is the projector onto the eigenspace of $M$ with eigenvalue $\lambda$, the probability of obtaining outcome $\lambda$ when measuring the density operator $\rho$ is $\trc{\rho P_\lambda}$, and the resulting state is 
	\[\rho'\ =\ \frac{P_\lambda\rho P_\lambda}{\trc{\rho P_\lambda}}.\]
\end{description}
Projective measurements are also known as von Neumann measurements. The Projective measurement postulate can be derived from the Measurement postulate. Indeed, using the fact that projectors are Hermitian, by the spectral theorem we have
\[\sum_{\lambda \in \spct{M}} \adj{P_\lambda}P_\lambda\ =\ \sum_{\lambda \in \spct{M}} {P_\lambda}^2\ =\ \sum_{\lambda \in \spct{M}} {P_\lambda}\ =\ \idty,\]
and by invariance of traces under cyclic permutations, we have:
\[\trc{\adj{P_\lambda}P_\lambda\rho}\ =\ \trc{P_\lambda^2\rho}\ =\ \trc{P_\lambda\rho}\ =\ \trc{\rho P_\lambda},\]
	so that
	\[\frac{P_\lambda\rho\adj{P_\lambda}}{\trc{\adj{P_\lambda}P_\lambda\rho}}\ =\ \frac{P_\lambda\rho P_\lambda}{\trc{\rho P_\lambda}}.\]

\newcommand{\measout}{\texttt{measure-outcome}}
\newcommand{\prjmeas}{\texttt{proj-measurement}}
\newcommand{\injon}{\texttt{inj-on}}
\newcommand{\moval}[1]{#1^\mathrm{v}}
\newcommand{\moprj}[1]{#1^\mathrm{p}}
\newcommand{\measopr}{\texttt{meas-outcome-prob}}

Projective measurements could be formalized in many ways. We chose to stick with a formalization that is as close as possible to the one used in \cite{liu19} for their quantum programs, for the sake of future reusability. We consider a measure outcome as a couple $(\lambda, P_\lambda)$, where $\lambda$ represents the output of the measure and $P_\lambda$ the associated projector, and introduce a binary predicate that characterizes  projective measurements. The first parameter of this argument represents the number of possible measure outcomes and the second parameter is the collection of measure outcomes. We require that the values of the measure outcomes are pairwise distinct, that the associated projectors have the correct dimensions and are orthogonal projectors, that sum to the identity. For the sake of readability, if $M_i = (\lambda, P_\lambda)$ is a measure outcome, then we denote $\lambda$ by $\moval{M_i}$ and $P_\lambda$ by $\moprj{M_i}$.

\[\begin{array}{l}
	\keyw{type-synonym}\ \measout\ =\ \dR\times \cpxmat 
\end{array}\]
\[\begin{array}{lcl}
	\prjmeas & :: & \dN \rightarrow (\dN \rightarrow \measout) \rightarrow \dB\lmspace\\
	\prjmeas\ n\ M\ & \Leftrightarrow & \injon\ (\lambda i.\ \moval{M_i})\ \ \interv{0}{n-1}\ \wedge\\
	& & \forall j < n.\, \moprj{M_j} \in \fcmats \wedge \projector\ \moprj{M_j}\ \wedge\\
	& & \forall i,j < n.\, i\neq j \Rightarrow \moprj{M_i}\cdot \moprj{M_j} = \vzero\ \wedge\\
	& & \sum_{j = 0}^{n-1} \moprj{M_j} = \idty	
\end{array}\]

According to the projective measurement predicate, the probability of obtaining result $\lambda$ when measuring the density operator $\rho$ is $\trc{\rho P_\lambda}$. We prove that, although $\rho$ and $P_\lambda$ are complex matrices, these traces are real positive numbers that sum to 1.

\[\begin{array}{lcl}
	\measopr & :: & \cpxmat \rightarrow (\dN \rightarrow \measout) \rightarrow \\
	& & \quad \quad \dN \rightarrow \dC\\
	\measopr\ \rho\ M\ i & = & \trc{\rho \cdot \moprj{M_i}}
\end{array}
\]

\[\begin{array}{l}
	\keyw{lemma}\ \textsc{meas-outcome-prob-real}:\\
	\aindent \keyw{assumes}\ \rho \in \fcmats\ \keyw{and}\ \densop\ \rho\\
	\aindent \keyw{and}\ \prjmeas\ n\ M\ \keyw{and}\ i < n\\
	\sindent \keyw{shows}\ \measopr\ \rho\ M\ i \in \dR\lmspace\\
	\keyw{lemma}\ \textsc{meas-outcome-prob-pos}:\\
	\aindent \keyw{assumes}\ \rho \in \fcmats\ \keyw{and}\ \densop\ \rho\\
	\aindent \keyw{and}\ \prjmeas\ n\ M\ \keyw{and}\ i < n\\
	\sindent \keyw{shows}\ \measopr\ \rho\ M\ i \geq 0\lmspace\\
	\keyw{lemma}\ \textsc{meas-outcome-prob-sum}:\\
	\aindent \keyw{assumes}\ \rho \in \fcmats\ \keyw{and}\ \densop\ \rho\\
	\aindent \keyw{and}\ \prjmeas\ n\ M\\
	\sindent \keyw{shows}\ \sum_{j = 1}^{n-1} (\measopr\ \rho\ M\ j) = 1
\end{array}\]

When the result of the measurement of $\rho$ is $\lambda$, $\rho$ collapses into $\frac{P_\lambda\rho P_\lambda}{\trc{\rho P_\lambda}}$. When formalizing this collapse in Isabelle, some care must be taken to handle the case of results that occur with probability zero. Although such cases are never meant to be analyzed when reasoning on quantum algorithms, it is still necessary to provide a reasonable definition of the state $\rho$ collapses into. We have chosen to make $\rho$ collapse into the maximally mixed state in this case:

\newcommand{\dtycol}{\texttt{density-collapse}}

\[\begin{array}{lcl}
	\dtycol & :: & \cpxmat \rightarrow \cpxmat \rightarrow \cpxmat\\
	\dtycol\ \rho\ P & = & \keyw{if}\ \trc{\rho \cdot P} = 0\ \keyw{then}\ \maxmix\ \dimr\\
	& & \quad\quad\quad\quad\quad\quad\quad\keyw{else}\ \frac{P\cdot\rho\cdot P}{\trc{\rho\cdot P}}
\end{array}
\]

\subsection{Projective measurements for observables}

\newcommand{\diagelems}{\texttt{diag-elems}}
\newcommand{\diagitoel}{\texttt{diag-idx-to-el}}
\newcommand{\diageli}{\texttt{diag-elem-indices}}
\newcommand{\prjvec}{\texttt{project-vecs}}
\newcommand{\mkmo}{\texttt{mk-meas-outcome}}
\newcommand{\eigvals}{\texttt{eigvals}}
\newcommand{\makepm}{\texttt{make-pm}}
\newcommand{\schur}{\texttt{unitary-schur-decomposition}}
\newcommand{\decard}{\texttt{dist-el-card}}

We develop the construction of a projective measurement for an observable, i.e., a Hermitian matrix. This construction relies on the fact that a Hermitian matrix $A$ can be decomposed as $A = U\cdot B\cdot \adj{U}$, where $B$ is a diagonal matrix and $U$ is unitary. A  construction of $B$ and $U$ based on the Schur decomposition theorem is available in Isabelle; this theorem was developed in \cite{jordan} and extended in \cite{liu19}. The projective measurement for $A$ is constructed using the fact that the spectrum of $A$ consists of the diagonal elements of $B$, and because $U$ is unitary, its column vectors are necessarily normalized and pairwise orthogonal. More specifically, assume $A\in \cmat{n}{n}$, let $D_B\isdef \setof{B_{i,i}}{i = 1, \ldots, n}$ (represented by $\diagelems\ B$ in our formalization), and let $p\isdef \card{D_B}$ (represented by $\decard\ B$ in our formalization). The number $p$ represents the size of the spectrum of $A$; we let $\calC$ (represented by $\diagitoel\ B$ in our formalization) be a bijection between $\interv{0}{p-1}$ and $D_B$, and $\calE$ (represented by $\diageli\ B$ in our formalization) associate to $\lambda \in D_B$ the set $\setof{i\leq n}{B_{i,i} = \lambda}$. Then the set $\set{\calE\circ\calC(0), \ldots, \calE\circ\calC(p-1)}$ is a partition of $\interv{1}{n}$ (lemmas \textsc{diag-elem-indices-disjoint} and \textsc{diag-elem-indices-union} in our formalization). This partition is used, along with the unitary matrix $U$, to construct projectors for the eigenspaces of $A$ as follows. We define the function $\prjvec$ which, for $i\in \interv{0}{p-1}$, constructs the matrix 
\[P_i\isdef \sum_{j\in \calE\circ\calC(i)} \outerp{U_j}{U_j},\]
where $U_j$ denotes column $j$ of $U$. For $i\in \interv{0}{p-1}$, the function $\mkmo\ i$ constructs the couple $(\calC(i), P_i)$. We obtain the definition of the projective measurement associated to the Hermitian matrix $A$ whose eigenvalues are represented by $\eigvals\ A$:
\[\begin{array}{lcl}
	\makepm & :: & \cpxmat \rightarrow \dN \times (\dN \rightarrow \measout)\\
	\makepm\ A & = & \keyw{let}\ (B,U,\_) = \schur\ A\ (\eigvals\ A)\\
	& & \keyw{in}\ (\decard\ B,\ \mkmo\ B\ U)
\end{array}
\]
The resulting couple represents a projective measurement, and the original matrix can be recovered by summing the projectors scaled by the corresponding eigenvalues:
\[\begin{array}{l}
	\keyw{lemma}\ \textsc{make-pm-proj-measurement}:\\
	\aindent \keyw{assumes}\ A \in \fcmats\ \keyw{and}\ \hermitian\ A\\
	\aindent \keyw{and}\ \makepm\ A = (n, M)\\
	\sindent \keyw{shows}\ \prjmeas\ n\ M\lmspace\\	
	\keyw{lemma}\ \textsc{make-pm-sum}:\\
	\aindent \keyw{assumes}\ A \in \fcmats\ \keyw{and}\ \hermitian\ A\\
	\aindent \keyw{and}\ \makepm\ A = (n, M)\\
	\sindent \keyw{shows}\ \sum_{i = 0}^{n-1} \moval{M_i}\cdot \moprj{M_i} = A
\end{array}\]

\section{The CHSH inequality}\label{sec:chsh}

\newcommand{\integrable}{\texttt{integrable}}
\newcommand{\jprob}[2]{\mathrm{p}(#1\mid #2)}
\newcommand{\jexp}[1]{\mathrm{E}(#1)}

The fact that a physical system is in a superposition of states and that, instead of revealing a pre-existing value, a measurement ``brings the outcome into being'' (Mermin, \cite{Mermin93}) was the cause of many controversies between the pioneers of quantum mechanics. Famously, Einstein did not believe in the intrinsically statistical nature of quantum mechanics. According to him, quantum mechanics was an incomplete theory, and the postulates on probabilistic measure outcomes actually reflected statistical outcomes of a deterministic underlying theory (see \cite{Dalton20,scarani2019bell} for detailed considerations on these controversies). The EPR paradox \cite{EPR} was designed to evidence the incompleteness of quantum mechanics. It involves two entangled and separated particles which are sent in opposite directions. If one of the particles is measured, then the outcome of the measurement of the other particle will be known with certainty. For example, if the entangled system is represented by the Bell state $\ket{\Phi^+} = \frac{1}{\sqrt{2}}\left(\ket{00} + \ket{11}\right)$ and a measurement of the first particle returns $0$, then it is certain that a measurement of the second one will also output $0$. This  phenomenon is known as \emph{nonlocality}, and it may leave the impression  that information traveled from the first to the second particle instantaneously, which would 
contradict the theory of relativity\footnote{Since then, it has been proven that this phenomenon is in no contradiction with the theory of relativity and does not imply faster-than-light communication.}. Einstein called this phenomenon ``spooky action at a distance''. A suggested solution to this phenomenon is that the measurement outcomes are actually properties that existed before the measurement was performed, and that deterministic underlying theories for quantum mechanics should thus be developed.
Efforts to develop such theories are called \emph{hidden-variable programs}. The theories that take into account the fact that information cannot travel instantaneously, thus also requiring that distant events are independent, are called \emph{local hidden-variable theories}. 

The fact that there can be no local hidden-variable underlying theory for quantum mechanics was proved by Bell \cite{Bell64} who derived inequalities (the \emph{Bell inequalities}), that hold in a probabilistic setting, and showed that they are violated by measurements in quantum mechanics. {Recently, Aspect \cite{aspect} showed that  this violation  can be experimentally verified, even when taking experimental errors into account,  {i.e.}, regardless of the possible outcomes of the particles lost during the experiment.}

In what follows we formalize probabilistic an inequality that is violated in quantum mechanics: the \emph{CHSH inequality}, named after Clause, Horne, Shimony and Holt \cite{CHSH}. This inequality is an upper-bound involving expectations of products of random variables. It involves two parties, Alice and Bob, who each receive and measure a particle that is part of an entangled system originating from a common source. After repeating this operation a large number of times, they can compute the frequencies of the different outcomes. These frequencies are referred to as \emph{correlations}, or \emph{joint probabilities}, and denoted by $\jprob{a,b}{A,B}$, where $a$ and $b$ represent the outcomes and $A$ and $B$ represent the measuring devices used by Alice and Bob, respectively. In the CHSH setting, Alice and Bob have two measuring devices each, represented by the observables $A_1$ and $A_2$ for Alice, and $B_1$ and $B_2$ for Bob. All observables have $\pm 1$ as eigenvalues.  At each round, Alice and Bob independently choose one measuring device; running the experimental sufficiently many times permits to construct the expectation values for observables $A_i$ and $B_j$, where $\set{i,j} \subseteq \set{1,2}$:
\[\jexp{A_i,B_j}\ \isdef\ \sum_{a,b = \pm 1} a\cdot b\cdot \jprob{a,b}{A_i,B_j}.\]
Consider the quantity 
\[\jexp{A_1,B_1} + \jexp{A_1,B_0} + \jexp{A_0, B_1} - \jexp{A_0, B_0}.\]
As we will see, under the local hidden-variable assumption, this quantity admits an upper-bound, but for a suitable choice of density operator and observables, this upper-bound is violated; a violation that has been confirmed experimentally \cite{aspect}.

Under the local hidden-variable assumption, the quantities $\jexp{A_i,B_j}$ are expectations in a suitable probability space. We prove the following inequality which holds in any probability space $\calM$, with relaxed conditions on the upper-bounds of random variables compared to standard statements of the result, which are assumed to hold almost everywhere rather than for all samples:
\[\begin{array}{l}
	\keyw{lemma}\ \textsc{chsh-expect}:\\
	\aindent \keyw{assumes}\ \AEv_\calM\,x.\ \card{A_0(x)} \leq 1\ \keyw{and}\ \AEv_\calM\,x.\ \card{A_1(x)} \leq 1\\
	\aindent \keyw{and}\ \AEv_\calM\,x.\ \card{B_0(x)} \leq 1\ \keyw{and}\ \AEv_\calM\,x.\ \card{B_1(x)} \leq 1\\
	\aindent \keyw{and}\ \integrable\ \calM\ (A_0\cdot B_0)\\
	\aindent \keyw{and}\ \integrable\ \calM\ (A_0\cdot B_1)\\
	\aindent \keyw{and}\ \integrable\ \calM\ (A_1\cdot B_0)\\
	\aindent \keyw{and}\ \integrable\ \calM\ (A_1\cdot B_1)\\
	\sindent \keyw{shows}\ \card{\dE[A_1\cdot B_0] + \dE[A_0\cdot B_1] + \dE[A_1\cdot B_1] - \dE[A_0\cdot B_0]} \leq 2	
\end{array}\]

The local hidden-variable assumption on a system states that there exists a probability space and independent random variables such that, when performing simultaneous measurements on the system, the probabilities of the outcomes, the values of which are given by the Measurement postulate, are expectations in the probability space. This statement is often found in articles and textbooks under the assumption that the probability space admits a density\footnote{Including in the original paper on the CHSH inequality \cite{CHSH}.}, but it is defined in our formalization in a more general case, where no assumption on the existence of a density is made, and properties on the considered random variables are assumed to hold almost everywhere rather than on the entire probability space.
\[\begin{array}{lcl}
	\posrv & :: & \alpha\ \msr \rightarrow (\alpha\rightarrow \dR) \rightarrow \dB\\
	\posrv\ \calM\ X & \equiv & X\in \bmeas\ \calM\ \wedge\ \AEv_\calM\, x.\  X(x) \geq 0\lmspace\\
	\psrv & :: & \alpha\ \msr \rightarrow \cpxmat \rightarrow (\dC\rightarrow \alpha \rightarrow \dR) \rightarrow \dB\\
	\psrv\ \calM\ A\ X & \equiv & \AEv_\calM\, x. \sum_{a\in \spct{A}} X_a(x) = 1\lmspace\\
	\lhv & :: & \alpha\ \msr \rightarrow \cpxmat \rightarrow\\
	& & \quad \cpxmat \rightarrow \cpxmat \rightarrow\\
	& & \quad  (\dC \rightarrow \alpha \rightarrow \dR) \rightarrow (\dC \rightarrow \alpha \rightarrow \dR) \rightarrow \dB\\
	\lhv\ \calM\ A\ B\ \rho\ X\ Y & \equiv & \pbspace \calM\ \wedge\\
	& & \psrv\ \calM\ A\ X\ \wedge\ \psrv\ \calM\ B\ Y\ \wedge\\
	& & \forall a\in \spct{A}.\, \posrv\ \calM\ X_a\ \wedge\\
	& & \forall b\in \spct{B}.\, \posrv\ \calM\ Y_b\ \wedge\\
	& & \forall a\in \spct{A}.\, \forall b\in \spct{B}.\\ 
	& & \quad (\integrable\ \calM\ (X_a\cdot Y_b)\ \wedge\\
	& & \quad \dE[X_a\cdot Y_b] = \trc{P_a\cdot P_b\cdot \rho})
\end{array}\]

\newcommand{\qtexp}{\texttt{qt-expect}}

The \emph{quantum expectation value} of a measurement represents the average value of the projective measurement of an observable. In other words, given an observable $A$, if the probability of obtaining result $a\in \spct{A}$ after a measurement of some state is $p_a$, then the expectation value of $A$ is $\sum_{a\in \spct{A}} a\cdot p_a$. More generally, given a density operator $\rho$ and an observable $A$, the (quantum) expectation value of $A$ is
\[\tuple{A}_\rho\ \isdef\ \trc{A\cdot\rho}.\]
Under the local hidden-variable hypothesis, when $X$ represents observable $A$, we can define the random variable related to the expectation value of $A$:

\[\begin{array}{lcl}
	\qtexp & :: & \cpxmat\rightarrow (\dC\rightarrow \alpha \rightarrow \dR) \rightarrow \alpha \rightarrow \dR\\
	\qtexp\ A\ X & = & \left(\lambda x.\, \sum_{a\in \spct{A}} a \cdot X_a(x)\right)
\end{array}\]

We obtain the following equality relating the expectation of the product of random variables and the quantum expectation of the corresponding observables:
\[\begin{array}{l}
	\keyw{lemma}\ \textsc{sum-qt-expect}:\\
	\aindent \keyw{assumes}\ \lhv\ \calM\ A\ B\ \rho\ X\ Y\\
	\aindent \keyw{and}\ A\in \fcmats\ \keyw{and}\ B\in \fcmats\ \keyw{and}\ \rho\in \fcmats\\
	\aindent \keyw{and}\ \hermitian\ A\ \keyw{and}\ \hermitian\ B\\	
	\sindent \keyw{shows}\ \dE[(\qtexp\ A\ X)\cdot(\qtexp\ B\ Y)] = \trc{A\cdot B\cdot \rho}
\end{array}\]

\newcommand{\tX}{\texttt{X}}
\newcommand{\tZ}{\texttt{Z}}
\newcommand{\tXpZ}{\texttt{XpZ}}
\newcommand{\tZmX}{\texttt{ZmX}}
\newcommand{\tXI}{\texttt{X-I}}
\newcommand{\tZI}{\texttt{Z-I}}
\newcommand{\tIXpZ}{\texttt{I-XpZ}}
\newcommand{\tIZmX}{\texttt{I-ZmX}}

The goal becomes finding a suitable density operator and suitable observables so that the combination of their traces violates the CHSH inequality. To this purpose, we consider the density operator 
\[\rho_C\ \isdef\ \outerp{\Psi^-}{\Psi^-},\ \text{where } \ket{\Psi^-}\ =\ \frac{1}{\sqrt{2}}\left(\ket{01} - \ket{10}\right) \text{ is one of the Bell states}.\]
We consider bipartite measurements of this entangled state. These measurements  involve the following observables:
\[\begin{array}{rclcrcl}
	\tZ & \isdef & \begin{pmatrix}
		1 & 0\\
	0 & -1 \end{pmatrix} & \quad & \tX & \isdef & \begin{pmatrix}
	0 & 1\\
	1 & 0
\end{pmatrix}\lmspace\\
	\tXpZ & \isdef & -\frac{1}{\sqrt{2}}(\tX + \tZ) & & \tZmX & \isdef & \frac{1}{\sqrt{2}}(\tZ - \tX)
\end{array}\]
These are all Hermitian matrices, and they are also unitary. The corresponding separated measurements are represented by the following tensor products:
\[\begin{array}{rclcrcl}
	\tZI & \isdef & \tZ\otimes \idty & \quad & \tXI & \isdef & \tX \otimes \idty\\
	\tIXpZ & \isdef & \idty \otimes \tXpZ &\quad & \tIZmX & \isdef & \idty \otimes \tZmX
\end{array}\]
Under the local hidden-variable hypothesis, we can compute the following expectations:
\[\begin{array}{l}
	\keyw{lemma}\ \textsc{Z-I-XpZ-chsh}:\\
	\aindent \keyw{assumes}\ \lhv\ \calM\ \tZI\ \tIXpZ\ \rho\ V_z\ V_p\\
	\sindent \keyw{shows}\ \dE[(\qtexp\ \tZI\ V_z)\cdot(\qtexp\ \tIXpZ\ V_p)] = \frac{1}{\sqrt{2}}\lmspace\\
	\keyw{lemma}\ \textsc{X-I-XpZ-chsh}:\\
	\aindent \keyw{assumes}\ \lhv\ \calM\ \tXI\ \tIXpZ\ \rho\ V_x\ V_p\\
	\sindent \keyw{shows}\ \dE[(\qtexp\ \tXI\ V_x)\cdot(\qtexp\ \tIXpZ\ V_p)] = \frac{1}{\sqrt{2}}\lmspace\\
	\keyw{lemma}\ \textsc{X-I-ZmX-chsh}:\\
	\aindent \keyw{assumes}\ \lhv\ \calM\ \tXI\ \tIZmX\ \rho\ V_x\ V_m\\
	\sindent \keyw{shows}\ \dE[(\qtexp\ \tXI\ V_x)\cdot(\qtexp\ \tIZmX\ V_m)] = \frac{1}{\sqrt{2}}\lmspace\\
	\keyw{lemma}\ \textsc{Z-I-ZmX-chsh}:\\
	\aindent \keyw{assumes}\ \lhv\ \calM\ \tZI\ \tIZmX\ \rho\ V_z\ V_m\\
	\sindent \keyw{shows}\ \dE[(\qtexp\ \tZI\ V_z)\cdot(\qtexp\ \tIZmX\ V_m)] = -\frac{1}{\sqrt{2}}	
\end{array}\]
Summing the first three expectation values and subtracting the last one returns $2\sqrt{2} > 2$, and the CHSH inequality is violated. We conclude that the local hidden-variable assumption cannot hold:

\[\begin{array}{l}
	\keyw{lemma}\ \textsc{no-lhv}:\\
	\aindent \keyw{assumes}\ \lhv\ \calM\ \tZI\ \tIXpZ\ \rho\ V_z\ V_p\\
	\aindent \keyw{and}\ \lhv\ \calM\ \tXI\ \tIXpZ\ \rho\ V_x\ V_p\\
	\aindent \keyw{and}\ \lhv\ \calM\ \tXI\ \tIZmX\ \rho\ V_x\ V_m\\
	\aindent \keyw{and}\ \lhv\ \calM\ \tZI\ \tIZmX\ \rho\ V_z\ V_m\\
	\sindent \keyw{shows}\ \text{False}	
\end{array}\]

\section{Conclusion}

%
%
%
We have formalized the essential notion of quantum projective measurements and the way they are obtained from observables. This formalization was carried out in a setting that is as general as possible. For example, the local hidden-variable hypothesis is formalized with as few conditions as possible and contrarily to many textbooks, makes no assumption on the existence of a density function on the underlying probability space.
We also took care of formalizing necessary notions in a way that should make them simple to reuse in other formalizations. For instance, the way projective measurements are defined is close to their usage in the quantum language used in \cite{liu19}, which could permit to consider extensions of this language in which it is possible to reason about measurement outcomes. This is a direction we are currently exploring: to the best of our knowledge, there are currently no formalized quantum languages that permit to perform probabilistic reasoning on quantum algorithms. Yet, this form of reasoning is common in textbooks on quantum mechanics, where it is obvious that if a large number of systems in the state $\frac{1}{\sqrt{2}}(\ket{0} + \ket{1})$ are measured in the standard basis, then approximately half of the outputs will be equal to 0; a simple consequence of the Law of large numbers which is formalized in Isabelle \cite{eberl}. Although the CHSH inequality shows that it is not possible to model quantum mechanics in a probabilistic setting, we are currently investigating how to associate a probability space to a quantum algorithm, in order to use the large corpus of results on measure theory that have already been formalized in Isabelle for the subsequent reasoning tasks.

The CHSH inequality turned out to be difficult to formalize, because although it is presented in a similar way in \cite{nielsen-book, mathQuantum}, we were unable to find a justification why their presentation entails the required result without making an additional assumption on the relationship between expectations of the random variables they consider and the quantum expectations of the related measurements. This is why the treatment of the CHSH inequality that is formalized in this paper is the one from \cite{scarani2019bell}, in which the local hidden-variable formulation is the same as in the original paper \cite{CHSH}. 
This inequality is also an important first step toward the formalization of \emph{device-independent} quantum cryptography protocols; i.e., protocols that are \emph{unconditionally secure}, even in the case where the devices used are noisy or malicious.  Indeed a key point of these protocols is the notion of \emph{self-testing}, which guarantees that 
 the verification of inequality violations with 
classical interaction is enough to certify quantum properties that are stronger than entanglement.

\paragraph{Acknowledgments.} The authors thank St\'ephane Attal for his feedback on the relationship between quantum and standard probabilities. 
This work benefited from the funding \emph{``Investissements d'avenir''} (ANR-15-IDEX-02) program of
the French National Research Agency.

\bibliography{biblio}

\begin{thebibliography}{10}

\bibitem{aspect}
A.~Aspect.
\newblock Experimental tests of {B}ell's inequalities in atomic physics.
\newblock In I.~Lindgren, A.~Ros{\'e}n, and S.~Svanberg, editors, {\em Atomic
  Physics 8}, pages 103--128, Boston, MA, 1983. Springer US.

\bibitem{Ballarin14}
C.~Ballarin.
\newblock Locales: {A} module system for mathematical theories.
\newblock {\em J. Autom. Reasoning}, 52(2):123--153, 2014.

\bibitem{Bell64}
J.~S. Bell.
\newblock {On the Einstein Podolsky Rosen paradox}.
\newblock {\em Physics}, 1(3):195--200, 1964.

\bibitem{Boender_2015}
J.~Boender, F.~Kamm\"uller, and R.~Nagarajan.
\newblock Formalization of quantum protocols using {C}oq.
\newblock {\em Electronic Proceedings in Theoretical Computer Science},
  195:71--83, Nov 2015.

\bibitem{isa-dirac}
A.~Bordg, H.~Lachnitt, and Y.~He.
\newblock Isabelle marries {D}irac: a library for quantum computation and
  quantum information.
\newblock {\em Journal of Automated Reasoning}, 2020.

\bibitem{CHSH}
J.~F. {Clauser}, M.~A. {Horne}, A.~{Shimony}, and R.~A. {Holt}.
\newblock {Proposed Experiment to Test Local Hidden-Variable Theories}.
\newblock {\em Phys. {R}ev. {L}ett.}, 23(15):880--884, Oct. 1969.

\bibitem{Dalton20}
B.~J. Dalton, B.~M. Garraway, and M.~D. Reid.
\newblock Tests for einstein-podolsky-rosen steering in two-mode systems of
  identical massive bosons.
\newblock {\em Phys. Rev. A}, 101:012117, Jan 2020.

\bibitem{Durrett}
R.~Durrett.
\newblock {\em Probability : theory and examples}.
\newblock The Wadsworth \& Brooks/Cole statistics/probability series. Wadsworth
  Inc. Duxbury Press, Belmont, California, 1991.

\bibitem{eberl}
M.~Eberl.
\newblock The laws of large numbers.
\newblock {\em Archive of Formal Proofs}, Feb. 2021.
\newblock \url{https://isa-afp.org/entries/Laws_of_Large_Numbers.html}, Formal
  proof development.

\bibitem{EPR}
A.~Einstein, B.~Podolsky, and N.~Rosen.
\newblock Can quantum-mechanical description of physical reality be considered
  complete?
\newblock {\em Phys. Rev.}, 47(10):777--780, May 1935.

\bibitem{linalg}
P.~A. Fuhrmann.
\newblock Linear systems and operators in hilbert space.
\newblock {\em Proceedings of the Edinburgh Mathematical Society},
  26(1):113--114, 1983.

\bibitem{hoelzl2012thesis}
J.~H{\"o}lzl.
\newblock {\em Construction and Stochastic Applications of Measure Spaces in
  Higher-Order Logic}.
\newblock PhD thesis, Institut f{\"u}r Informatik, Technische Universit{\"a}t
  M{\"u}nchen, October 2012.

\bibitem{Kammuller19}
F.~Kamm{\"{u}}ller.
\newblock Attack trees in {I}sabelle extended with probabilities for quantum
  cryptography.
\newblock {\em Comput. Secur.}, 87, 2019.

\bibitem{types-sets}
O.~Kuncar and A.~Popescu.
\newblock From types to sets by local type definition in higher-order logic.
\newblock {\em J. Autom. Reason.}, 62(2):237--260, 2019.

\bibitem{liu19}
J.~Liu, B.~Zhan, S.~Wang, S.~Ying, T.~Liu, Y.~Li, M.~Ying, and N.~Zhan.
\newblock Formal verification of quantum algorithms using {Q}uantum {H}oare
  {L}ogic.
\newblock In I.~Dillig and S.~Tasiran, editors, {\em Computer Aided
  Verification}, pages 187--207. Springer International Publishing, 2019.

\bibitem{Mermin93}
N.~D. Mermin.
\newblock Hidden variables and the two theorems of john bell.
\newblock {\em Rev. Mod. Phys.}, 65:803--815, Jul 1993.

\bibitem{nielsen-book}
M.~A. Nielsen and I.~L. Chuang.
\newblock {\em Quantum Computation and Quantum Information: 10th Anniversary
  Edition}.
\newblock Cambridge University Press, USA, 10th edition, 2011.

\bibitem{ConcreteSemantics}
T.~Nipkow and G.~Klein.
\newblock {\em Concrete Semantics: With Isabelle/HOL}.
\newblock Springer Publishing Company, Incorporated, 2014.

\bibitem{Coq_Quantum}
R.~Rand, J.~Paykin, and S.~Zdancewic.
\newblock Qwire practice: Formal verification of quantum circuits in {C}oq.
\newblock {\em Electronic Proceedings in Theoretical Computer Science},
  266:119--132, 02 2018.

\bibitem{scarani2019bell}
V.~Scarani.
\newblock {\em Bell Nonlocality}.
\newblock Oxford Graduate Texts. Oxford University Press, 2019.

\bibitem{mathQuantum}
W.~Scherer.
\newblock {\em Mathematics of Quantum Computing: An Introduction}.
\newblock Springer International Publishing, 01 2019.

\bibitem{jordan}
R.~Thiemann and A.~Yamada.
\newblock Formalizing jordan normal forms in isabelle/hol.
\newblock In J.~Avigad and A.~Chlipala, editors, {\em Proceedings of the 5th
  {ACM} {SIGPLAN} Conference on Certified Programs and Proofs, Saint
  Petersburg, FL, USA, January 20-22, 2016}, pages 88--99. {ACM}, 2016.

\bibitem{Unruh_2019}
D.~Unruh.
\newblock Quantum relational hoare logic.
\newblock {\em Proceedings of the ACM on Programming Languages}, 3(POPL):1--31,
  Jan 2019.

\end{thebibliography}
\bibliographystyle{abbrv}
\end{document}